\documentclass[submission,copyright,creativecommons]{eptcs}
\providecommand{\event}{VORTEX2018} 

\usepackage{breakurl}             
\usepackage[english]{babel}
\usepackage{underscore}           
\usepackage{graphicx}
\usepackage{subcaption}

\usepackage{amsmath}
\usepackage{amssymb}

\usepackage[utf8]{inputenc}
\usepackage[T1]{fontenc}

\usepackage{latexsym}

\usepackage[utf8]{inputenc}
\usepackage{listings}
\usepackage{xspace}
\usepackage{hyperref}

\usepackage{cleveref}

\usepackage{xcolor}
\usepackage{pgfplots}
\usepackage{tikz}

\usepackage[shortlabels]{enumitem}
\usepackage{calc}

\usepackage{doi}

\newcommand\eg{e.g.,\xspace}
\newcommand\cf{c.f.,\xspace}
\newcommand\ie{i.e.,\xspace}

\newcommand\rw[1]{[\![#1]\!]}
\newcommand\lub\sqcup
\newcommand\pc{\texttt{pc}}
\newcommand\pcl{\texttt{pc\%$\ell$}}
\newcommand\pcd{\texttt{pc\%$d$}}
\newcommand\df{\triangleq}
\newcommand\dfsep{&\df&}

\newcommand\this{\mathit{\textsf{this}}}
\newcommand\goto{\mathit{\textsf{goto}}}
\newcommand\op{\mathit{op}}
\newcommand\new{\mathit{\textsf{new}}}
\newcommand\ifff{\mathit{\textsf{if}}}
\newcommand\return{\mathit{\textsf{return}}}

\newcommand\aloe{\mathit{\textsf{assert}}}
\newcommand\rname[1]{\textsc{#1}}

\title{SNITCH: Dynamic Dependent Information Flow Analysis for Independent Java Bytecode}
\author{Eduardo Geraldo \quad João Costa Seco
\institute{NOVA LINCS - Faculdade de Ciências e Tecnologia da Universidade Nova de Lisboa\\Portugal}
\and
}
\def\titlerunning{SNITCH: Dynamic Information Flow Analysis for Independent Java Bytecode}
\def\authorrunning{E. Geraldo \& J.C. Seco}

\begin{document}
\maketitle

\begin{abstract}

Software testing is the most commonly used technique in the industry to certify the correctness of software systems. This includes security properties like access control and data confidentiality.
However, information flow control and the detection of information leaks using tests is a demanding task without the use of specialized monitoring and assessment tools.

In this paper, we tackle the challenge of dynamically tracking information flow in third-party Java-based applications using dependent information flow control. Dependent security labels increase the expressiveness of traditional information flow control techniques by allowing to parametrize labels with context-related information and allowing for the specification of more detailed and fine-grained policies.
Instead of the fixed security lattice used in traditional approaches that defines a fixed set of security compartments, dependent security labels allow for a dynamic lattice that can be extended at runtime, allowing for new security compartments to be defined using context values.

We present a specification and instrumentation approach for rewriting JVM compiled code with in-lined reference monitors. 
To illustrate the proposed approach we use an example and a working prototype, SNITCH. SNITCH operates over the static single assignment language Shimple, an intermediate representation for Java bytecode used in the SOOT framework.



\end{abstract}
\
\section{Introduction}\label{sec:intro}

Data confidentiality is central in current software engineering practices.
In the past years, there have been recurrent news about information leaks surfacing as a result of subtle programming errors.
For instance, GitHub\footnote{http://bit.ly/2XNfEEU}
and
Twitter\footnote{http://bit.ly/2XuMH16},
both large-scale systems with impact on a large number of users, discovered and reported that their users' passwords were stored in cleartext to internal system logs, from where an ill-intended employee could have access to them and enter the users' accounts.
Certifying the functional correctness of software systems by testing is commonly accepted as a satisfactory approximation for compliance with functional specifications and requirement fulfilment in the software industry. However, testing aspects such as data confidentiality is a difficult task when using traditional approaches.
Properties like access control and information flow control require setting up complex testing scenarios where the symptoms of an error are hardly detectable. Typically, information leaks are only perceived at a global scale by detailed observation of side-effects. 



Information flow analysis~\cite{Austin2009,Denning1976,Sabelfeld2003,Zdancewic2004} is a language-based approach for information leak detection on software systems. Information flow analysis is present in the literature, in the form of static and dynamic analysis, each with their advantages and disadvantages.
Static analysis usually requires a considerable effort in code annotation or the complete refactoring of the target system. Besides, the over-approximation of results and the existence of false positives also poses an obstacle to the usability of languages and tools~\cite{tooljoana2013atps,JIF,FlowCaml} that employ this kind of analysis.
The alternative presented by dynamic analysis techniques produces less false positives but has some problems of its own.
It needs exhaustive testing to achieve maximum coverage, and it is subject to label creeping~\cite{AlmeidaMatos2009,Sabelfeld2003}, \ie the monotonic increase of the security label of values rendering them unusable.

Information flow control mechanisms depend on a security lattice~\cite{Denning1976} whose security labels are, usually, fixed. Consequently, existing are usually too restrictive when defining information flow policies, only allowing to define coarse-grained security policies which are inadequate in many cases. For instance, one often groups all users of a system under the same security label ``User'', which does not prevent user $A$ from accessing information of a user $B$.
We follow a more expressive approach that introduces dependent types for information flow~\cite{Ferreira2012,Lourenco2015} and access-control~\cite{Caires2011} allowing for the definition of data-dependent policies.
Value-dependent security labels improve the expressiveness of traditional information flow techniques. By allowing the parametrization of security labels with context-related information, it is possible not only to define more detailed and fine-grained policies, but also to create new security compartments at runtime.
Java Information Flow (JIF)~\cite{Myers2003, JIF} supports dynamic labels which differ from dependent security labels. Dynamic labels follow a decentralized security model~\cite{Myers2003} based on the notion of data ownership and authorizations. According to this model, each data item has an owner, and the owner allows, or not, its data to be read or written by some entity.
Dependent security labels follow a traditional security model with a security lattice that hierarchically organizes security labels and where a datum has a security label and can only be accessed by entities with sufficient privileges.
For instance, using dependent security labels, we can define policies restricting access to an employee's personal telephone number to the employee itself and its department manager.

In this paper, we present a strategy to specify and rewrite the intermediate Java code of applications to embed reference monitors capable of enforcing information flow policies using dependent security labels.
Our low-level code rewriting approach for the Java virtual machine language is inspired by tools like SASI~\cite{erlingsson2000} and TaintDroid~\cite{Enck2014} that automatically monitor the confidentiality of information of compiled Java programs.
To check data confidentiality in an application, TaintDroid~\cite{Enck2014} instruments the underlying runtime system (Android Java virtual machine) while our approach instruments the application itself.
We require the specification of some selected classes that make up the entry points of a system, for instance, service controllers~\cite{Daigneau2011} or DAO classes\footnote{Database access objects. Dtatypes that match database table schemas.}~\cite{Alur2003}. 
Then, we use this information to introduce in-lined reference monitors and instrument the application code to taint computed values with dependent security labels.
Our approach follows the style and semantics of the seminal work by Austin and Flanagan on dynamic information flow analysis~\cite{Austin2009} and is inspired by works such as the one by Lourenço and Caires~\cite{Lourenco2015,Lourenco2016} on dependent information flow analysis and the work of Chandra and Franz~\cite{Chandra2007} about hybrid information flow analysis for Java bytecode.

The rewriting process, presented in \autoref{sec:approach} operates over an intermediate representation in the static single assignment form~\cite{Appel1998,Zhao2013} (SSA). SSA is a way to arrange operations such that each variable is defined only once and allows to simplify and improve some optimizations such as constant propagation, value numbering, common sub-expression elimination and partial redundancy elimination among others.
We present an example that illustrates the rewriting process over an intermediate representation in the SSA form.
Our approach is backed by a prototype tool, SNITCH, to instrument intermediate Java code. SNITCH was evaluated on small-scale web applications using Java servlets and it uses the SOOT framework\footnote{https://github.com/Sable/soot}~\cite{Vallee-Rai1999} for code rewriting, a framework for optimizing and manipulating Java bytecode and offers multiple intermediate representations. One of such representations is Shimple, an intermediate representation in the SSA form over which our prototype operates. 

Our contributions can be summarized as follows:

\begin{itemize}
  \item a rewriting system for instrumenting static single assignment instructions with in-line reference monitors for information flow control with dependent information flow labels;

  \item a specification schema to define dependent information flow policies;

  \item a way to define dependent security labels in Java; and

  \item a tool capable of instrumenting third-party compiled code with an in-line reference monitor.
\end{itemize}

We leave for future work the introduction of abstract interpretation to optimize the computation of security labels and mechanisms like the one presented by Austin and Flanagan~\cite{Austin2010} to reduce label creeping and increase the number of accepted programs. 
Abstract interpretation would allow us not only to reduce the number of security label related computations executed at runtime, but also to achieve a gradual approach.

We start this paper by briefly presenting some concepts on dependent information flow labels in \Cref{sec:dif}. \Cref{sec:approach} presents our approach and describes an example of a web application. As we present our approach we also describe the steps required to instrument the example application to test it for information leaks. In \Cref{sec:examplell} we illustrate the code rewriting process. In Sections \ref{sec:valid} and \ref{sec:rw} we provide validate our approach and discuss the related work. Finally, in \Cref{sec:conclusions}, we conclude with some remarks on how to pursue this line of work.


\section{Dependent Security Information Flow}\label{sec:dif}

Language-based security~\cite{LBS2001}, and in particular information flow control~\cite{Denning1976}, specify and provide a platform to enforce security policies from the perspective of data creation, manipulation and data flow operations.
Information flow control allows the definition of hierarchic security compartments and the tracking of all uses of data, ensuring that higher security data does not flow (leak) to unrelated or lower security compartments.
Traditionally, Security labels are organized in lattices~\cite{Denning1977} and are associated with value types at compile-time~\cite{Myers2003,ZdancewicPHD} or used to taint values at runtime~\cite{Austin2009,Chandra2007}.

Information flow control allows for the detection of both explicit and implicit illegal information flows.
Explicit flows result from data transfer operations such as assignments, while implicit information flows arise from the control flow of a program.
High security label computations can have side effects on values of lower security labels allowing those with access to lower labelled variables to infer the values of those computations.
The side effects of high security computations on lower labelled values go against the non-interference property, a property at the core of information flow control and that denotes the absence of information leaks.
According to non-interference, changes to high security label values must not reflect themselves on lower security labelled values, \ie changes in the secret input of a program must not interfere with the program's public output~\cite{Zdancewic2004}.



Traditional security lattices enforce a significant degree of label squashing due to the lack of precision of the security labels used.
For instance, it is usually the case that a single security label is used to represent all the users of a system, not allowing to define fine-grained, per user, information flow restrictions.
The introduction of dependent security policies increases the preciseness of security specifications and introduces a higher degree of flexibility and usability. 
Dependent security policies are present in approaches like value-dependent information flow types~\cite{Ferreira2012,Lourenco2015,Lourenco2016}, and dynamic labels~\cite{JIF}.
The former was first introduced in the context of access control policies in~\cite{Caires2011} and then extended to the domain of static checking of information flow in~\cite{Ferreira2012,Lourenco2015,Lourenco2016}.
With dependent security labels, we can express, for instance, that a given function yields values of a parametric security label \verb!user(u)! where \verb!u! is a runtime value, allowing for row-level compartmentalization of security data visibility (\cf~\cite{Lourenco2015}).
The predefined lattice is automatically extended to capture dependent security labels like \verb!user(u)!, \verb!user(!$\top$\verb!)!, and \verb!user(!$\bot$\verb!)!.
The generic security label \verb!user(!$\top$\verb!)! is the label taht allows access to all users' data. 
The security label \verb!user(!$\bot$\verb!)! is the label all users can read.
We have the relations \verb!user(!$\bot$\verb!)! $\sqsubset$ \verb!user(u)! $\sqsubset$ \verb!user(!$\top$\verb!)!, for any user \verb!u!, and \verb!user(u)! $\#$ \verb!user(v)! for all users \verb!u! and \verb!v! such that \verb!u! $\not=$ \verb!v!. The first relation means that, for any user \verb!u!, data can flow from the user's security compartment \verb!user(u)!, to label \verb!user(!$\top$\verb!)!, and from \verb!user(!$\bot$\verb!)! to label \verb!user(u)!.


\section{Technical Approach}\label{sec:approach}

\begin{figure}[tb]
  \centering
  \includegraphics[width=.8\linewidth]{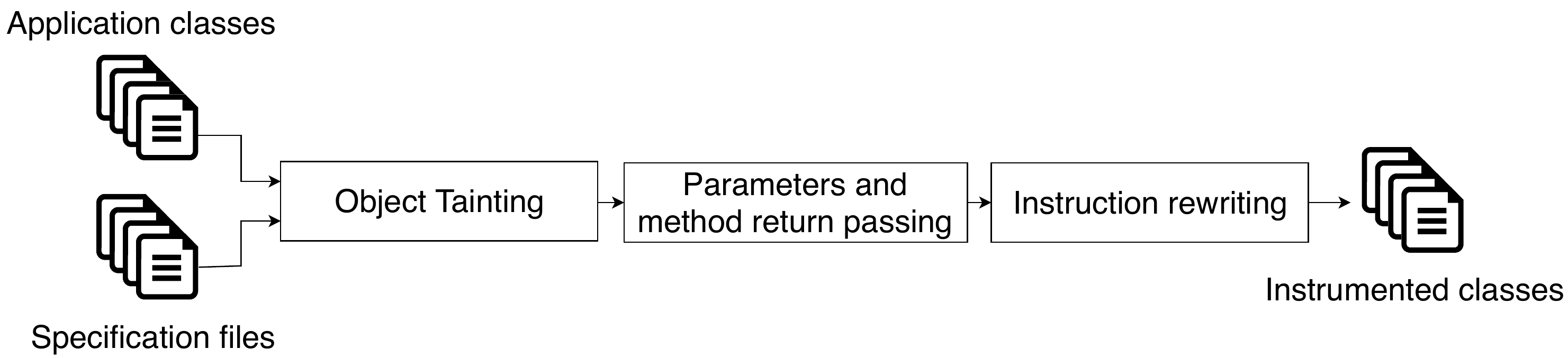}
  \caption{Code Instrumentation Phases}
  \label{fig:architecture}
\end{figure}

Our technical approach follows the phases depicted in \autoref{fig:architecture}. 
Given an application, a dependent security lattice, and a security specification for key classes of the application, our approach instruments the application with an in-lined reference monitor capable of enforcing information flow policies.
We next illustrate our approach with the help of an example.

Let us consider a small web application to implement a directory for a given company integrated in their website.
The information stored by such a system for each employee includes its identifier, name, address, salary, and its password.
We define two kinds of employees in this example: supervisor and associate. The latter category includes extra information, namely its supervisor and information about its last evaluation. Other users include unregistered users accessing the company's website.

In this example, we consider the following access constraints to the stored information:
\begin{itemize}
  \item only registered users can see the address of other users (employees);

  \item an associate employee can only access its salary;

  \item a supervisor user can access all associate users' salary information;

  \item the information regarding who supervises who can only be accessed by supervisor users;

  \item the information about the evaluation of an associate user can only be accessed by supervisor users;

  \item passwords are always secret, and no one but its owner should access it.
\end{itemize}

\noindent
Our example implements both retrieval and insertion operations. The operations are the following: it is possible to list the employees in the system; to retrieve the information about a specific employee; to compute the average salary; to add new employees to the system.
Only a registered user, an employee, can execute the operation that retrieves the information about any other employee. This operation exhibits different behaviours depending on who is executing it and what information is retrieved.

\begin{figure}[t]
    \centering
  \includegraphics[height=0.2\paperheight]{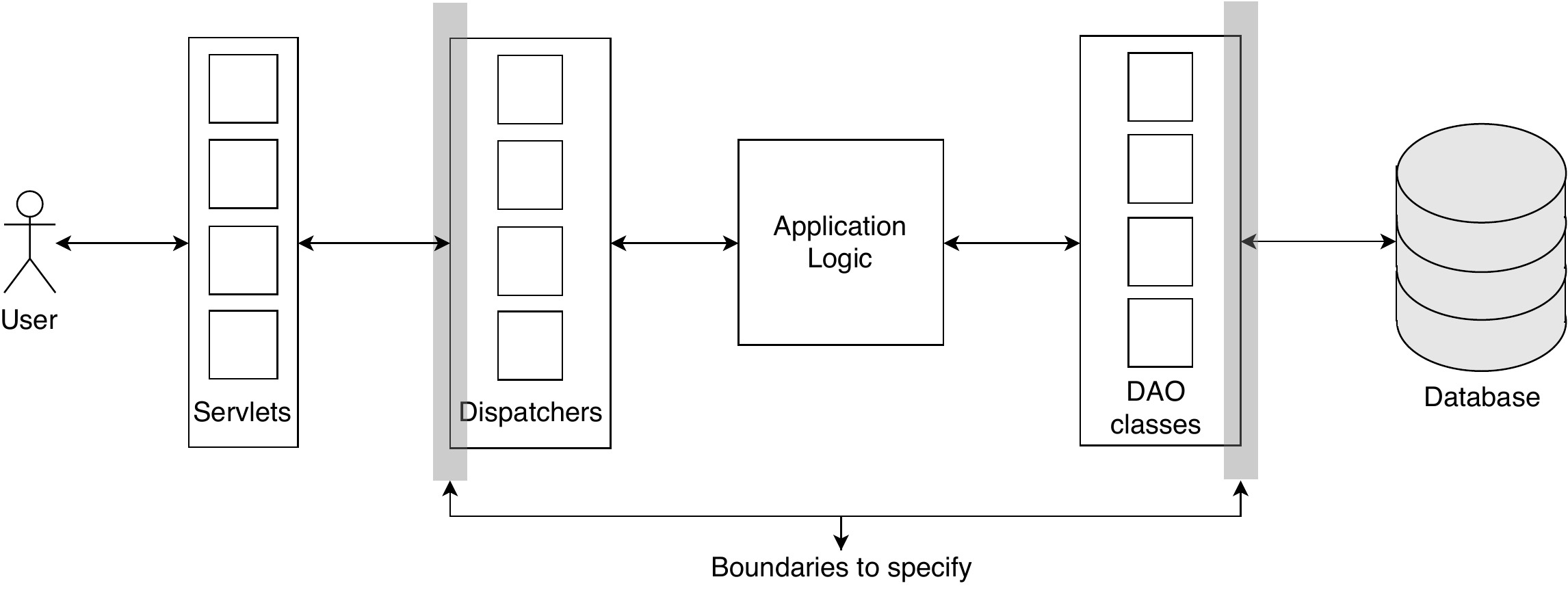}
  \caption{Typical architecture of a web application}
    \label{fig:app_arch}      
\end{figure}
In \autoref{fig:app_arch} we illustrate the generalised architecture of a web application.
For the sake of space and simplicity, in our evaluation example we merged the application logic and request dispatchers in a single layer.
In \autoref{fig:dao_classes} we present the DAO classes to represent the company's employees. Class \texttt{Associate} represents associate employees and Class \texttt{Supervisor} represents supervisor employees. 
Both classes, \texttt{Associate} and \texttt{Supervisor}, extend the abstract class \texttt{Employee} that contains the information common to both kinds of employees.
Class \texttt{Supervisor} is empty since it does not add any new fields to class \texttt{Employee}.
We omit all class methods as there are only getters and setters. 

\paragraph{Dependent Security Labels} The instrumentation of a system starts by defining the security specification. 
First, it is necessary to define and implement custom security labels to extend the default security lattice provided by SNITCH which only includes two built-in labels. The label \texttt{Public}, the lowest security label of all, and the label \texttt{Secret}, the highest security label of all.

Custom security labels are implemented by extending the abstract class \texttt{SecurityLevel}, provided in a companion library, and by defining a required comparator method. 
By implementing the missing comparator method, we model the security lattice used by the reference monitor during the system's execution.
Each label requires also a constructor which takes the same parameters as the security label. For each label parameter, the label's contructor requires one extra parameter of type \texttt{int}. 
For instance, a custom label \texttt{User}, parametrized by a \texttt{String} and a \texttt{long} has a constructor with the signature \texttt{User(String,int,long,int)}.
The extra parameter tells the monitor if the value passed to the security label's constructor is $\bot$, $\top$, or if the corresponding parameter is to be considered as is.

\begin{figure}[htbp]
    \centering
  \includegraphics[height=0.3\paperheight]{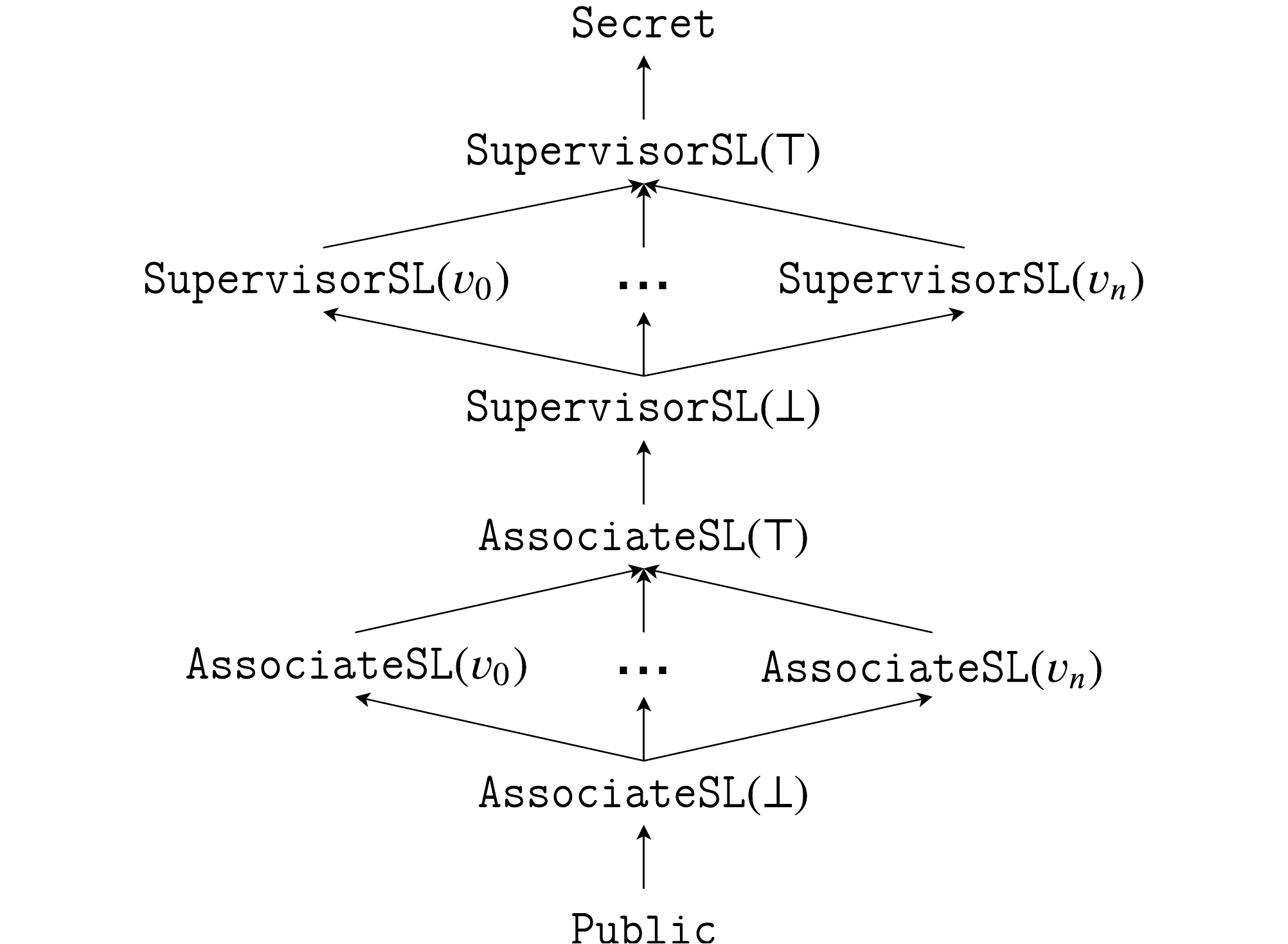}
  \caption{Security lattice used to instrument example.}
    \label{fig:app_lattice}      
\end{figure}
In order to instrument the example, we define two custom security labels, \texttt{SupervisorSL}, and \texttt{AssociateSL}, which define security compartments for supervisors and associates, respectively. 
Both labels have a single parameter, an employee id, and their comparator methods define the security lattice depicted in \autoref{fig:app_lattice}.


\begin{figure}[tb]
  \centering
\begin{minipage}[t]{0.5\linewidth}
        \centering
\begin{lstlisting}[frame=none, numbers=none, xleftmargin=.2\textwidth, xrightmargin=.3\textwidth, escapeinside={(*@}{@*)}]
abstract class Employee {
    long id;
    String name;
    String address;
    double salary;
    String pwd;
    (*@{\ldots}@*)
}
\end{lstlisting}
        \label{fig:dao_employee}
    \end{minipage}%
    \begin{minipage}[t]{0.5\linewidth}
        \centering
\begin{lstlisting}[frame=none, numbers=none, xleftmargin=.2\textwidth, escapeinside={(*@}{@*)}]
class Associate extends Employee {
    Supervisor supervisor;
    double evaluation;
    (*@{\ldots}@*)
}

class Supervisor extends Employee {}
\end{lstlisting}
        \label{fig:dao_associate}
    \end{minipage}%
    \vspace{0pt}
\caption{Classes \texttt{Employee}, \texttt{Associate}, and \texttt{Supervisor}}
  \label{fig:dao_classes}
\end{figure}

\paragraph{Specifications files}
Besides custom security labels, it is necessary to define a security layer, using a set of specification files and the security lattice (custom security labels) used to in-line the reference monitor.
In \autoref{fig:app_arch} we show the layer that needs specification, the classes that make the boundaries of the system. 
This layer includes all the classes that communicate with the exterior context of the system, such as service controllers and DAO classes in a typical architecture.

Specifications files include annotations to define the security labels of class fields and method signatures.
The semantics of field annotations is the following. If a security label is explicitly assigned to a class field, it will be fixed (as maximum) throughout the entire execution.
Any attempt to store a value in such a field will result in one of two outcomes:
if the incoming value's security label is lower than the expected security label, then the incoming value's security label is upgraded;
if the incoming value's security label is higher than the expected field's security label, the monitor signals an information leak.
When a field is not annotated with a security label, it changes according to the stored values.
We, however, do not let assignment operations to lower the security labels of  variables or fields (\cf~\cite{Austin2009}) to avoid implicit information leaks.

When defining specifications for methods, it is possible to annotate both parameters and return values. Method annotations differ from field annotations as they include one of two modifiers, \texttt{?} or \texttt{!}. 
If an annotation uses the modifier \texttt{?}, the reference monitor compares the security label of the annotated value with the security label used in the annotation and, if higher, the monitor signals an information leak.
If an annotation uses the modifier \texttt{!}, the monitor will associate the annotation's security label with the annotated value.
The modifier \texttt{!} allows one to associate a security label with input from outside the system and to declassify information.
Since it allows for information declassification~\cite{AlmeidaMatos2009}, it is necessary to use this modifier carefully as it may easily lead to incorrect specifications which result in undetected information leaks.
If a parameter or return value does not have a security annotation, then the monitor will propagate its security label without performing any extra operation.

In the current example, we define specification files for all dispatcher classes and for classes that represent stored information about employees.
The specifications for classes \texttt{Employee}, \texttt{Associate}, and \texttt{Supervisor} are depicted in \autoref{fig:spec_dao_employee}, \autoref{fig:spec_dao_supervisor}, and \autoref{fig:spec_dao_associate} respectively. Notice that all fields in the classes are annotated with a security label from the lattice in \autoref{fig:app_lattice}. Security label parameters are instantiated with fields to denote a concrete dependency, or \verb!(_)! to represent $\bot$. 
The values used to instantiate security label parameters must belong to the same object and produce a security dependency between fields. For instance, in class \texttt{Supervisor} we use \texttt{SupervisorSL(id)} as a security label dependent on the value of field \texttt{id}, declared in the superclass \texttt{Employee}. The security label \texttt{AssociateSL(\_)} establishes a security compartment accessible to all \texttt{Associate} employees.
%

\begin{figure}[tb]
  \centering
\begin{minipage}[t]{0.5\linewidth}
        \centering
\begin{lstlisting}[frame=none, numbers=none, escapeinside={(*@}{@*)}]
abstrcat class Employee {
    long:Public id;
    String:Public name;
    String:AssociateSL(_) address;
    String:Secret pwd;   
}
\end{lstlisting}
\subcaption{}
\label{fig:spec_dao_employee}
\end{minipage}%
\begin{minipage}[t]{0.5\linewidth}
\centering
\begin{lstlisting}[frame=none, numbers=none, escapeinside={(*@}{@*)}]
class Supervisor extends Employee {
    double:SupervisorSL(id) salary;
}
\end{lstlisting}
\subcaption{}
\label{fig:spec_dao_supervisor}
\end{minipage}
\begin{minipage}[b]{0.5\linewidth}
\centering
\begin{lstlisting}[frame=none, numbers=none, escapeinside={(*@}{@*)}]
class Associate extends Employee {
    double:AssociateSL(id) salary;
    Supervisor:SupervisorSL(_) supervisor;
    double:SupervisorSL(supervisor) evaluation;   
}
\end{lstlisting}
\subcaption{}
\label{fig:spec_dao_associate}
\end{minipage}%
\caption{Security specifications for classes \texttt{Employee}, \texttt{Supervisor} and \texttt{Associate}}
\label{fig:dao_specs}
\end{figure}


\begin{figure}[t]
  \centering
\begin{lstlisting}[showstringspaces=false,frame=none, numbers=none, xleftmargin=.15\textwidth, xrightmargin=.3\textwidth, escapeinside={(*@}{@*)}]
public String associateDispatch(long requesterId, long queriedId) {

   Employee queried = EmployeeRepository.getInstance().getEmployeeById(queriedId);

   String response = "{"
      + "\"id\": " + "\"" + queried.getId() + "\"" 
      + "\"name\": \"" + queried.getName() + "\","
      + "\"address\": \"" + queried.getAddress() + "\",";

   if (requesterId == queriedId)
      response += "\"salary\": \"" + queried.getSalary() + "\",";
        
   return response + "}";
}
\end{lstlisting}
\caption{Source code for the employee information dispatcher}
\label{fig:dispatcher_class}
\end{figure}

\begin{figure}[t]
\centering
\begin{lstlisting}[frame=none, numbers=none, xleftmargin=0\textwidth, escapeinside={(*@}{@*)}]
class EmployeeInfoDispatcher {
   String:?AssociateSL(requesterId) associateDispatch(long requesterId, long queriedId);
}
\end{lstlisting}
\caption{Security specifications for the employee information dispatcher}
\label{fig:dispatcher_spec}
\end{figure}

The effort required to write security specifications depends on several factors: the knowledge about the system to instrument, the constraints of the system, and the complexity of the security specification and lattice.
After defining the security specifications and labels, it is possible to instrument the application.
\paragraph{Value tainting}
In the first instrumentation step, we inject shadow fields in every application class. One of them holds the security label of the class instance while the remaining ones mirror existing class fields of a primitive or library (non-instrumented) type.

\paragraph{Methods and Parameter passing}
In order to propagate the security labels of primitive or library-type arguments and return values, we add shadow fields to each class. These shadow fields help preparing method calls by allowing the caller to store and the callee to retrieve the arguments' security labels. When the called method terminates its execution, the callee stores the security labels of the arguments and return value for the caller to retrieve.

\paragraph{Instruction rewriting}
The final step of the instrumentation process consists in the instrumentation of method bodies.
The body of a method consists of a graph of basic blocks, where a basic block is a sequence of instructions starting with a label and terminating in a return, a branching, or a jump instruction.
The instrumentation process compositionally rewrites instructions in the SSA form~\cite{Appel1998,Zhao2013}, according to the rules defined in \autoref{fig:rules}. Every rule for instructions that give place to information flows take into account the security label associated with the computation itself, \ie the security label of the program counter ($\pcl$)~\cite{Austin2009}.

The set of instructions considered is the following: load a value to a local variable \underline{$v = s$} or \underline{$v=o.f$}; method call \underline{$v = o.g(v_1...v_n)$}, where \underline{$o$} represents the target object; object instantiation \underline{$v = \new \; C$}; binary operations \underline{$v = \op(e,e)$}; phi expressions \underline{$v = \phi(v,v)$}; conditional jumps \underline{$\ifff(e)\;\goto\;l$}; unconditional jumps \underline{$\goto\:l$}; and the return instruction \underline{$\return\;e$}.
A phi function is a pseudo-function used in merge points to yield one of its arguments according to the control-flow path executed.


\begin{figure}[tb]
	\centering
$\begin{array}{r@{\,}c@{\,}lll}
\rw {v := k}_\ell \dfsep v := k;\; v_s := v_s \lub \pcl && \rname{(const)}\\[0.2em]

\rw {v := v'}_\ell \dfsep v := v';\; v_s := v_s \lub v'_s \lub \pcl && \rname{(local)}\\[0.2em]

\rw {v := f}_\ell \dfsep v := f;\; v_s := v_s \lub f_s \lub \pcl && \rname{(field)}\\[0.2em]

\rw {f := v}_\ell \dfsep \aloe(v_s \lub \pcl \sqsubseteq f_s);\; f := v; & \text{($f$ has spec.)} & \rname{(fieldC)}\\[0.2em]

\rw {f := v}_\ell \dfsep f := v;\; f_s := f_s \lub v_s \lub \pcl & \text{($f$ has no spec.)} & \rname{(fieldW)}\\[0.2em]

\rw {v := \new \; C}_\ell \dfsep v := \new\; C;\; v_s := v_s \lub \pcl && \rname{(new)}\\[0.2em]

\rw{v := o.g(v_1, \ldots, v_n)}_\ell \dfsep \multicolumn{2}{@{}l}{o.f_{p1} := v_{1_s}\lub \pcl; \ldots; o.f_{pn} := v_{n_s}\lub \pcl;}\\
&& v := o.g(v_1, \dots, v_n);\\
&& v_{1_s} := o.f_{p1}; \ldots; v_{n_s} := o.f_{pn};\\
&& v_s := o.f_{\mathit{return}} & \text{($o$ has spec.)} & \rname{(call)}\\[0.2em]

\rw {v := o.g(v_1,\ldots,v_n)}_\ell \dfsep v := o.g(v_1,\ldots,v_n);\\
&& v_s := v_s \lub o_s \lub v_{1_s} \lub \ldots\lub v_{n_s} \lub \pcl & \text{($o$ has no spec.)} & \rname{(callX)}\\[0.2em]

\rw {v := \op(e_0, e_1)}_\ell \dfsep v := \op(e_0, e_1);\\
&& v_s := v_{0_s} \lub v_{1_s} \lub \pcl && \rname{(bin op)}\\[0.2em]

\rw {\ifff (v) \;\goto \; k}_\ell \dfsep \pcl_{out} := \pcl_{in} \lub v_s \\ && \ifff (v) \;\goto\:k && \rname{(branch)}\\[0.2em]

\rw {\goto\:k}_\ell \dfsep \goto\:k && \rname{(goto)}\\[0.2em]

\rw {v := \phi(v_0, v_1)}_\ell \dfsep v =: \phi(v_0, v_1); v_s := v_s \lub \phi(v_{0_s}, v_{1_s}) && \rname{(phi)}\\[0.2em]

\rw {\return\;e}_\ell \dfsep 

\this.f_{\return} := e_s \lub \pcl;\\[0.2em]
&&o.f_{p1} := v_{1_s}; \ldots; o.f_{pn} := v_{n_s};\\[0.2em]
&&\return\;e; && \rname{(return)}\\[0.2em]
\end{array}
$
	\caption{Instrumentation rules}
	\label{fig:rules}
\end{figure}

\noindent
\emph{Notation:} we use \underline{$C$} to denote class names, \underline{$v$} and \underline{$o$} to denote local variables or registers, \underline{$k$} to denote value literals, \underline{$e$} ranges over local variables and constants; \underline{$f$} to denote object fields, and \underline{$g$} to denote method names. A variable \underline{$v_s$} stores the security label of variable \underline{$v$}, a field \underline{$f_s$} stores the security label of field \underline{$f_s$}, a field \underline{$g_{pi}$} stores the security label of parameter \underline{$i$} of the method $g$ and a field \underline{$g_{\mathit{ret}}$} stores the security label of the return value of method $g$.


\begin{figure}[tb]
\centering
\begin{minipage}[t]{0.5\linewidth}
\centering
$
\begin{array}{rl}
\ell: & \pcl := \phi(\pcl_{1},...,\pcl_{n})\\
& \rw{e_1}\\
& ...\\
& \rw{e_n}\\[0.5em]
& \ell_1,...,\ell_n \text{ are predecessor nodes of $\ell$.}
\end{array}
$
\subcaption{}
\label{fig:context_merge}
\end{minipage}%
\begin{minipage}[t]{0.5\linewidth}
        \centering
$
\begin{array}{rl}
\ell: & \pcl_{in} := \pcd_{in}\\
& \rw{e_1}\\
& ...\\
& \rw{e_n}\\[0.5em]
& \mathit{d} \text{ is the post-dominated node}\\
& \text{where the current scope started.}
\end{array}
$
\subcaption{}
\label{fig:context_pop}
\end{minipage}%

$$
\pcl_{out} \text{ is equal to }\pcl_{in} \text{ if not stated otherwise.}
$$
\caption{Context security label rules for merging control flows}
\label{fig:pc_rules}
\end{figure}

The rules for loading operations, depicted in \autoref{fig:rules}, (\rname{const}, \rname{local}, and \rname{field}) work by combining the $\pc$, the security label of the value, and the variable's security label.
The dynamic modification of a field, rule \rname{fieldW}, potentially increases the field's security label with combination of $\pc$ and the value's security label.
Rule \rname{fieldC} applies to fixed label fields, where writes are always lower or equal to the current label. 

To deal with a method call, we have two rules.
Rule \rname{call} handles the call to an instrumented method. It starts by copying the arguments' security labels to the target method with the help of auxiliary fields and then calls the method. Once the method completes its execution, we retrieve the result and argument's security label. It is necessary to collect the argument security labels after the call since they might have changed during the method's execution.
Rule \rname{callX} accounts for the use of non-instrumented methods, where the resulting security label is the combination of all operands' security labels plus the program counter and, in the case of instance methods, the callee's security label.

Notice that in the case of rule \rname{branch}, the value for the context's security label ($\pc$) increases according to the security label of the branch condition.
Once the execution leaves the scope started by the branch condition, it is necessary to reinstate $\pc$'s old security label. To restore the $\pc$, we follow the rules depicted in \autoref{fig:pc_rules}.
The rule depicted in \autoref{fig:context_merge} is applied in the case where there are multiple predecessors ($\ell_1,...,\ell_n$) of the basic block $\ell$ but $e_1$ does not post-dominate a common predecessor of $\ell_1,...,\ell_n$, \ie multiple control flows converge but the scope does not change. According to this rule, the security label of the context at beginning of the basic $\ell$ results from the $\phi$ function of the predecessors' context security labels.
The second rule, the rule depicted in \autoref{fig:context_pop}, applies when there are multiple predecessors ($\ell_1,...,\ell_n$) of $\ell$  and $e_1$ is the first instruction to post-dominate a common predecessor, $d$, of $\ell_1,...,\ell_n$. In this case, the context security label at beginning of block $\ell$ ($\pcl_{in}$) is equal to the context security label at the beginning of $d$ ($\pcd_{in}$), \ie $e_1$ is the first instruction to execute outside the scope created in $d$ and reinstates the value of $\pc$ before entering the new scope.
Unconditional branches do not change any security meta-information. Rule~\rname{Phi} chooses the security label according to the executed predecessor.
When a return instruction executes, the $\pc$ stack is placed at the same label as it was when the function was called (because of ad-hoc returns at any point in the method body). Besides restoring the $\pc$, it is also necessary to copy the returned value's security label and the argument's security labels to auxiliary fields.
It is necessary to update the security labels of the arguments to deal with cases where they are objects of non-instrumented types. The objects' security label might change during the method's execution, in which case it is necessary to propagate any changes to the caller.

\paragraph{Testing phase} 
Once defined the security layer and instrumented the application, we test the example for information leaks. 
To do so, we need to test all available operations in the instrumented application.
If an operation has an information leak, the monitor halts the system's execution indicating an assertion violation.
Strong guarantees about data confidentiality depend on the test coverage achieved.

In summary, our approach for the detection of illegal information flows using dependent security labels, is embodied in a tool based on the SOOT framework which instruments a target application with a reference monitor. With this approach, we believe to have improved the process of security certification for third-party systems. Despite the need for some specification effort, typically, there is a set of DAO and controller classes that are known and for which is possible to design a specification.

\section{Rewriting Process Example}\label{sec:examplell}


\begin{figure}[tb]
  \centering
\begin{minipage}[t]{0.5\linewidth}
        \centering
\begin{lstlisting}[frame=none, numbers=none, xleftmargin=.3\textwidth, xrightmargin=.3\textwidth, escapeinside={(*@}{@*)}]
class Example {
  SecurityLabel secLbl$this;

  SecurityLabel secLbl$field0;
  long field0;

  Example field1;
    
  SecurityLabel secLbl$methodA$p0;
  SecurityLabel secLbl$methodA$p1;
  int methodA (int a, int b) {(*@{\ldots}@*)}
  SecurityLabel secLbl$methodA$ret;

  Example methodB(Example e) {(*@{\ldots}@*)}

}
\end{lstlisting}
\subcaption{Class \texttt{Example} after field injection.}
  \label{fig:ll_class_int}
\end{minipage}%
\begin{minipage}[t]{0.5\linewidth}
\centering
\begin{lstlisting}[frame=none, numbers=none, xleftmargin=.3\textwidth, xrightmargin=.3\textwidth, escapeinside={(*@}{@*)}]
class Example {

  long field0;
  Example field1;
    
  int methodA (int a, int b) {
    if(a > b)
      return a;
    return b;
  }

  Example methodB(Example e) {(*@{\ldots}@*)}
  
}
\end{lstlisting}
\subcaption{Method \texttt{methodA} of Class \texttt{Example}.}
\label{fig:class_ll_init}
\end{minipage}
\caption{Class \texttt{Example}}
\label{fig:unused_label}
\end{figure}

In this section we illustrate the rewriting process using a small example.
Let us consider the Java class depicted in \autoref{fig:class_ll_init}.
As stated in \autoref{sec:approach}, first, we add new fields (to which we will refer as ``label fields'') to the application classes for storing security labels. 
We add one label field for the objects' security label (\texttt{secLbl\$this}), one label field for every field of a non-instrumented type (\texttt{secLbl\$field0}) and for every method we add label fields for parameters and return values of non-instrumented types (\texttt{secLbl\$methodA\$p0}, \texttt{secLbl\$methodA\$p1}, and \texttt{secLbl\$methodA\$ret}).
We show the result the of field injection on class \texttt{Example} in \autoref{fig:ll_class_int}.



\begin{figure}[tb]
  \centering
  \begin{minipage}[t]{0.5\linewidth}
        \centering
\begin{lstlisting}[frame=none, numbers=left, xleftmargin=.11\textwidth, escapeinside={(*@}{@*)}]
class Example {

  (*@{\ldots}@*)
    
  int methodA (int a, int b) {
    if(a > b) goto LABEL0
    result_0 = b 
    goto LABEL1
LABEL0:
    result_1 = a;
LABEL1:
    result_2 = phi(result_0, result_1)
    return result_2
  }

  (*@{\ldots}@*)

}
\end{lstlisting}
    \subcaption{}
    \label{fig:method_ssa_orig}
  \end{minipage}%
  \begin{minipage}[t]{0.5\linewidth}
    \centering
\begin{lstlisting}[frame=none, numbers=left, escapeinside={(*@}{@*)}]
class Example {

  (*@{\ldots}@*)
    
  int methodA (int a, int b) {
    secLbl$a = this.secLbl$methodA$p0
    secLbl$b = this.secLbl$methodA$p1
    secLbl$cond = combine(secLbl$a, secLbl$b)
    secLbl$oldPC = increasePC(cond)
    if(a > b) goto LABEL0
    result_0 = b
    secLbl$result_0 = secLbl$b
    goto LABEL1
LABEL0:
    result_1 = a;
    secLbl$result_1 = secLbl$a
LABEL1:
    setPC(secLbl$oldPC)
    result_2 = phi(result_0, result_1)
    secLbl$result_2
       = phi(secLbl$result_0, secLbl$result_1)
    this.secLbl$methodA$ret = secLbl$result_2
    this.secLbl$methodA$p0 = secLbl$a
    this.secLbl$methodA$p1 = secLbl$b
    return result_2
  }

  (*@{\ldots}@*)

}
\end{lstlisting}
    \subcaption{}
    \label{fig:method_ssa_instr}
  \end{minipage}%
\caption{Original (left) and instrumented (right) SSA code for method \texttt{methodA} of Class \texttt{Example}}
  \label{fig:method_ssa_both}
\end{figure}

Once injected all the necessary fields, we can proceed to rewrite the methods' body in the SSA form. To do so, we rewrite every instruction according to the rules defined in \Cref{sec:approach}.
\autoref{fig:method_ssa_orig} depicts a possible representation of \texttt{methodA} in the static single assignment form and, \autoref{fig:method_ssa_instr} illustrates the result of \texttt{methodA}'s body rewriting.

\begin{enumerate}[leftmargin=\widthof{[[Lines_XX-YY]}+\labelsep]
  \item[Lines 6-7] retrieve arguments' security labels (\texttt{secLlb\$a} and \texttt{secLlb\$b}) from auxiliary fields injected for the purpose (\texttt{secLlb\$methodA\$p0} and \texttt{secLlb\$methodA\$p1} respectively);

  \item[Lines 8-10] compute the condition's security label. Then, update $\pc$, keeping its old value so that we can restore it when the execution leaves the branch's scope. Finally execute the branching instruction.

  \item[Lines 11-16] compute the value to return, security operations accompany every operation executed.
  Each branch stores the result in a different version of the same variable (\texttt{result_1} and \texttt{result_2}).

  \item[Lines 17-19] terminate the context initiated with the branching instruction (more specifically in line 9). When the execution leaves the scope of the branch instruction, it is necessary to restore $\pc$ to its previous value. Since there are two paths converging, it is also necessary to decide which version of the variables to consider using $\phi$ functions.

  \item[Lines 19-20] conclude the method's execution. They store the result's label (\texttt{secLbl\$result_2}) in field \texttt{secLbl$methodA$ret} and update the arguments' label fields. The return  only executes after storing the labels.
\end{enumerate}

\section{Experimental Validation}\label{sec:valid}

To provide some validation to our approach, we developed a prototype tool, SNITCH, and the instrumented the web application presented to provide some validation to our approach, we developed a prototype tool, SNITCH, and then used it to instrument the web application presented in \Cref{sec:approach}.
\begin{figure}[htbp]
    \centering
  \includegraphics[height=0.2\paperheight]{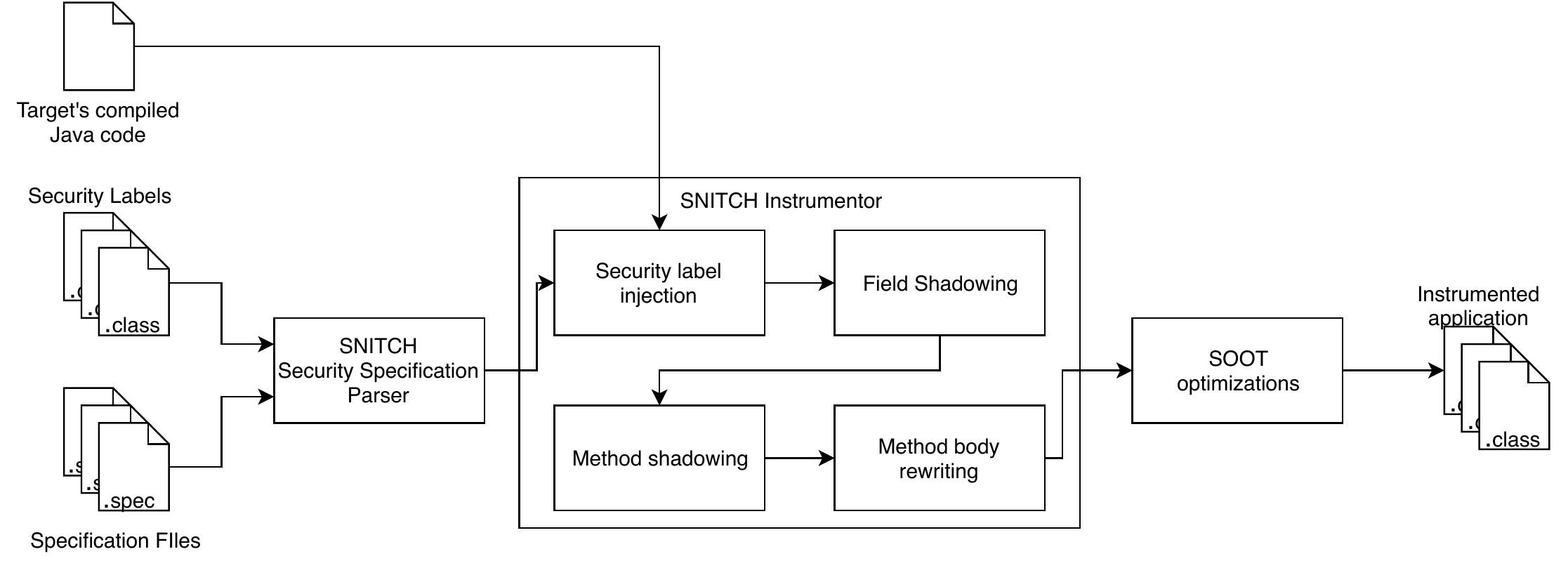}
  \caption{SNITCH internal phases}
    \label{fig:proto_arch}      
\end{figure}

Just as defined in the approach; SNITCH, based on a set of security specifications, instruments a system with an in-lined reference monitor. As can be seen \autoref{fig:proto_arch}, which depicts SNITCH's architecture,  SNITCH consists of two modules; a parser for the security specifications and an instrumentation module for bytecode rewriting. The latter component makes use of the SOOT~\cite{Vallee-Rai1999} framework which, as previously stated, is a framework for Java bytecode manipulation and optimization.
In an attempt to reduce the reference monitor's impact on the system's execution, SNITCH makes use of the optimization suites SOOT offers for optimizing the instrumented code.

To test the approach, we introduced information leaks in the example application. The leaks resulted from implicit and explicit information flows.
The leaks caused by explicit flows were bad assignments or attempts to return classified information.
The in-lined reference monitor in the application was capable of detecting all the information leaks introduced in the example application.

The example's information retrieval methods' implementation was naive, returning all information available on the employees disregarding any information access restrictions. We reached the final implementation of the application through a trial and error process in which we instrumented, tested, and fixed the application multiple times until no further information leaks were detected.

%
\definecolor{lgray}{HTML}{D8D8D8}
\definecolor{dgray}{HTML}{6E6E6E}
\definecolor{bblack}{HTML}{000000}
\definecolor{wwhite}{HTML}{FFFFFF}
%
\pgfplotsset{compat=1.11,
    /pgfplots/ybar legend/.style={
    /pgfplots/legend image code/.code={%
       \draw[##1,/tikz/.cd,yshift=-0.25em]
        (0cm,0cm) rectangle (3pt,0.8em);},
   },
}
\begin{figure}[htbp]
    \centering
  \begin{tikzpicture}
    \begin{axis}[
        width  = 0.75*\textwidth,
        height = 5cm,
        major x tick style = transparent,
        ybar=2*\pgflinewidth,
        bar width=14pt,
        ymajorgrids = true,
        ylabel = {Overhead factor (x1)},
        xlabel = {Application method},
        xlabel near ticks,
        symbolic x coords={List Employees, Get Employee Info, Get Avg Salary, Add new Supervisor, Add New Associate},
        xtick = data,
        x tick label style={font=\footnotesize,align=right,rotate=30},
        scaled y ticks = false,
        enlarge x limits=0.25,
        ymin=0,
        legend cell align=left,
        legend style={
                at={(1,1.05)},
                anchor=south east,
                column sep=1ex
        }
    ]

        \addplot[style={bblack,fill=lgray,mark=none}]
            coordinates {(List Employees,2.34) (Get Employee Info,2.75) (Get Avg Salary,2.02) (Add new Supervisor,1.58) (Add New Associate,1.40)};

    \end{axis}
\end{tikzpicture}
  \caption{Runtime overhead per application method.}
    \label{fig:runtime_chart}
\end{figure}
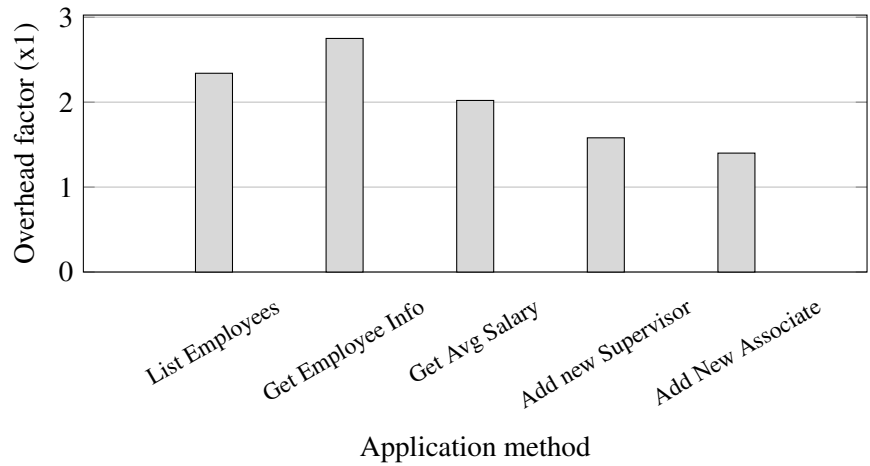

The instrumentation of the example web application also the collection of for the collection of some broad measurements on the reference monitor's impact on the execution time of an application. Still, more applications need to be instrumented and tested to obtain more accurate values.
We defined a set of five operations which we used to measure the total execution time and each operation's average execution time. We measured operations' execution time in both the original and instrumented applications.
\autoref{fig:runtime_chart} shows how the reference monitor affects the execution time of each operation. The operation for information retrieval is the one where the impact of the monitor was the greatest. An explanation for this is that this operation extracts the most information per employee; therefore, it executes data combination operations (\eg string concatenation) requiring more intervention from the monitor.
The overall execution time of the five hundred operations in the instrumented application is 1.79 times the original execution time. However, this estimate does not take into account the overhead of input and output operations which takes milliseconds to complete while CPU operations take microseconds.

From the instrumentation of the example, it was possible to observe the impact of the reference monitor on a system's execution time. Despite that the instrumented application is of a small dimension, we observed a significant overhead (average 1.79x) on the application's CPU time; however, there is still room for improvement; for instance, the prototype does not apply any kind of optimization specific to information flow control, only using the standard optimizations supplied by the SOOT framework.

\section{Related Work}\label{sec:rw}

There is a considerable amount of work on information flow analysis in the literature, ranging from axiomatic approaches~\cite{Andrews1980}, dynamic analyses~\cite{Austin2009,erlingsson2000}, programming languages and types~\cite{tooljoana2013atps,JIF, FlowCaml} to instrumented virtual machines~\cite{Enck2014}. Java Information Flow~\cite{Myers2003, JIF}, contains embedded information flow analysis capabilities, and allows the definition of a form of dynamic matching between labels and principals which can in turn be used to parametrize classes and define richer runtime policies.

TaintDroid~\cite{Enck2014} is an approach that does not extend or create a programming language but instruments the virtual machine where the intermediate language executes.
Sensitive data is tainted at its source (\eg GPS) and the instrumented virtual machine propagates the taint along a program's execution. When tainted data reaches a sink (\eg network interface), the information leak is logged. An advantage of the approach taken by TaintDroid over JIF is that it is not necessary to change application code.

Austin and Flanagan~\cite{Austin2009} present a dynamic approach for information flow analysis that guarantees non-interference in dynamically-typed languages. It presents and compares two approaches. \emph{Universal Labelling}, where all values have an explicit label (security label); and \emph{Sparse Labelling} where all values are tracked but only some are explicitly labelled. Sparse labelling is  observably equivalent to universal labelling but with significantly less overhead.

Ferreira~\cite{Ferreira2012} introduces the use of refinement types in information flow analysis. It presents an extension of the LiveWeb/$\lambda_{DB}$~\cite{Caires2011} with type-based information flow. Security labels are expressed using first-order logic propositions dependent on runtime values.
Value-dependent security labels are further developed by \cite{Lourenco2015}, who presents the first non-interference result for dependent information flow types.


\section{Concluding Remarks}\label{sec:conclusions}

The purpose of this work is to study the applicability of information flow analysis to the certification of third-party Java-based software systems. 
To convey a more usable, flexible and expressive framework, we have adopted dependent information flow control as the preferred abstraction.


This paper presents work in the development of a certification tool that attaches in-lined reference monitors to existing compiled code, based on interface specifications in observable points of systems.
We foresee some immediate follow-ups on this work, the challenges in dealing with label creeping and the introduction of abstract interpretation to help reduce the runtime overhead beyond the optimizations resulting from the use of SSA intermediate language.
Considering the security label combination operation ($\lub$) that given two security labels $\ell_A$ and $\ell_B$, yields the lowest security label that is higher or equal to both $\ell_A$ and $\ell_B$ and the security lattice shown in \autoref{fig:app_lattice}, computations such as $AssociateSL(\bot) \lub AssociateSL(\top)$ can be removed as its result is known beforehand ($AssociateSL(\top)$ ). By statically analysing the code, not only trivial computations could be removed, but also, it would be possible, in some instances, to detect illegal information flows statically. The introduction of such mechanisms would allow our approach to evolve from a dynamic information flow control mechanism to a hybrid one.
Another possible line of work would be the introduction of our approach in software development frameworks as a development tool. This would allow software developers to test their code as they develop. There also some advantages that our approach can benefit from if integrated with software development tools like the automatic extraction of specifications based on the frameworks annotations. Frameworks like Spring and Jersey annotate classes with information relevant to the specification files; for instance spring uses the annotation \texttt{@Entity} to flag DAO classes.

Regarding the monitor's overhead presented on \Cref{sec:valid}, we would like to highlight that the measurements made only took into account CPU time. When taking into account I/O operations, we can consider the monitor's overhead as negligible. For instance, the measurements of CPU time were of the order of the microseconds, while I/O operations took milliseconds, three decimal orders of magnitude greater and network operations even worse.

{\footnotesize
\paragraph{Acknowledgements} This work was funded by NOVA LINCS UID/CEC/04516/2013, COST CA15123 - FC\&T Project CLAY - PTDC/EEI-CTP/4293/2014}

\nocite{*}
\bibliographystyle{eptcs}
\bibliography{bibfile}

\begin{thebibliography}{10}
\providecommand{\bibitemdeclare}[2]{}
\providecommand{\surnamestart}{}
\providecommand{\surnameend}{}
\providecommand{\urlprefix}{Available at }
\providecommand{\url}[1]{\texttt{#1}}
\providecommand{\href}[2]{\texttt{#2}}
\providecommand{\urlalt}[2]{\href{#1}{#2}}
\providecommand{\doi}[1]{doi:\urlalt{http://dx.doi.org/#1}{#1}}
\providecommand{\bibinfo}[2]{#2}

\bibitemdeclare{article}{AlmeidaMatos2009}
\bibitem{AlmeidaMatos2009}
\bibinfo{author}{Ana \surnamestart Almeida~Matos\surnameend} \&
  \bibinfo{author}{G{\'e}rard \surnamestart Boudol\surnameend}
  (\bibinfo{year}{2009}): \emph{\bibinfo{title}{On Declassification and the
  Non-disclosure Policy}}.
\newblock {\sl \bibinfo{journal}{J. Comput. Secur.}}
  \bibinfo{volume}{17}(\bibinfo{number}{5}), pp. \bibinfo{pages}{549--597},
  \doi{10.3233/JCS-2009-0355}.

\bibitemdeclare{book}{Alur2003}
\bibitem{Alur2003}
\bibinfo{author}{Deepak \surnamestart Alur\surnameend}, \bibinfo{author}{Dan
  \surnamestart Malks\surnameend}, \bibinfo{author}{John \surnamestart
  Crupi\surnameend}, \bibinfo{author}{Grady \surnamestart Booch\surnameend} \&
  \bibinfo{author}{Martin \surnamestart Fowler\surnameend}
  (\bibinfo{year}{2003}): \emph{\bibinfo{title}{Core J2EE Patterns (Core Design
  Series): Best Practices and Design Strategies}}, \bibinfo{edition}{2}
  edition.
\newblock \bibinfo{publisher}{Sun Microsystems, Inc.},
  \bibinfo{address}{Mountain View, CA, USA}.

\bibitemdeclare{article}{Andrews1980}
\bibitem{Andrews1980}
\bibinfo{author}{Gregory~R. \surnamestart Andrews\surnameend} \&
  \bibinfo{author}{Richard~P. \surnamestart Reitman\surnameend}
  (\bibinfo{year}{1980}): \emph{\bibinfo{title}{An Axiomatic Approach to
  Information Flow in Programs}}.
\newblock {\sl \bibinfo{journal}{ACM Trans. Program. Lang. Syst.}}
  \bibinfo{volume}{2}(\bibinfo{number}{1}), pp. \bibinfo{pages}{56--76},
  \doi{10.1145/357084.357088}.

\bibitemdeclare{article}{Appel1998}
\bibitem{Appel1998}
\bibinfo{author}{Andrew~W. \surnamestart Appel\surnameend}
  (\bibinfo{year}{1998}): \emph{\bibinfo{title}{SSA is Functional
  Programming}}.
\newblock {\sl \bibinfo{journal}{SIGPLAN Not.}}
  \bibinfo{volume}{33}(\bibinfo{number}{4}), pp. \bibinfo{pages}{17--20},
  \doi{10.1145/278283.278285}.

\bibitemdeclare{inproceedings}{Austin2009}
\bibitem{Austin2009}
\bibinfo{author}{Thomas~H. \surnamestart Austin\surnameend} \&
  \bibinfo{author}{Cormac \surnamestart Flanagan\surnameend}
  (\bibinfo{year}{2009}): \emph{\bibinfo{title}{Efficient Purely-dynamic
  Information Flow Analysis}}.
\newblock In: {\sl \bibinfo{booktitle}{Proceedings of the ACM SIGPLAN Fourth
  Workshop on Programming Languages and Analysis for Security}},
  \bibinfo{series}{PLAS '09}, \bibinfo{publisher}{ACM}, \bibinfo{address}{New
  York, NY, USA}, pp. \bibinfo{pages}{113--124}, \doi{10.1145/1554339.1554353}.

\bibitemdeclare{inproceedings}{Austin2010}
\bibitem{Austin2010}
\bibinfo{author}{Thomas~H. \surnamestart Austin\surnameend} \&
  \bibinfo{author}{Cormac \surnamestart Flanagan\surnameend}
  (\bibinfo{year}{2010}): \emph{\bibinfo{title}{Permissive Dynamic Information
  Flow Analysis}}.
\newblock In: {\sl \bibinfo{booktitle}{Proceedings of the 5th ACM SIGPLAN
  Workshop on Programming Languages and Analysis for Security}}, pp.
  \bibinfo{pages}{3:1--3:12}, \doi{10.1145/1814217.1814220}.

\bibitemdeclare{inproceedings}{Caires2011}
\bibitem{Caires2011}
\bibinfo{author}{Lu\'{\i}s \surnamestart Caires\surnameend},
  \bibinfo{author}{Jorge~A. \surnamestart P{\'e}rez\surnameend},
  \bibinfo{author}{Jo\~{a}o~Costa \surnamestart Seco\surnameend},
  \bibinfo{author}{Hugo~Torres \surnamestart Vieira\surnameend} \&
  \bibinfo{author}{L\'{u}cio \surnamestart Ferr\~{a}o\surnameend}
  (\bibinfo{year}{2011}): \emph{\bibinfo{title}{Type-based Access Control in
  Data-centric Systems}}.
\newblock In: {\sl \bibinfo{booktitle}{Proceedings of the 20th European
  Conference on Programming Languages and Systems: Part of the Joint European
  Conferences on Theory and Practice of Software}},
  \bibinfo{series}{ESOP'11/ETAPS'11}, \bibinfo{publisher}{Springer-Verlag},
  \bibinfo{address}{Berlin, Heidelberg}, pp. \bibinfo{pages}{136--155},
  \doi{10.1006/inco.1994.1093}.

\bibitemdeclare{inproceedings}{Chandra2007}
\bibitem{Chandra2007}
\bibinfo{author}{D.~\surnamestart Chandra\surnameend} \&
  \bibinfo{author}{M.~\surnamestart Franz\surnameend} (\bibinfo{year}{2007}):
  \emph{\bibinfo{title}{Fine-Grained Information Flow Analysis and Enforcement
  in a Java Virtual Machine}}.
\newblock In: {\sl \bibinfo{booktitle}{Twenty-Third Annual Computer Security
  Applications Conference (ACSAC 2007)}}, pp. \bibinfo{pages}{463--475},
  \doi{10.1109/ACSAC.2007.37}.

\bibitemdeclare{book}{Daigneau2011}
\bibitem{Daigneau2011}
\bibinfo{author}{Robert \surnamestart Daigneau\surnameend}
  (\bibinfo{year}{2011}): \emph{\bibinfo{title}{Service Design Patterns:
  Fundamental Design Solutions for SOAP/WSDL and RESTful Web Services}},
  \bibinfo{edition}{1} edition.
\newblock \bibinfo{publisher}{Addison-Wesley Professional}.

\bibitemdeclare{article}{Denning1976}
\bibitem{Denning1976}
\bibinfo{author}{Dorothy~E. \surnamestart Denning\surnameend}
  (\bibinfo{year}{1976}): \emph{\bibinfo{title}{A Lattice Model of Secure
  Information Flow}}.
\newblock {\sl \bibinfo{journal}{Commun. ACM}}
  \bibinfo{volume}{19}(\bibinfo{number}{5}), pp. \bibinfo{pages}{236--243},
  \doi{10.1145/360051.360056}.

\bibitemdeclare{article}{Denning1977}
\bibitem{Denning1977}
\bibinfo{author}{Dorothy~E. \surnamestart Denning\surnameend} \&
  \bibinfo{author}{Peter~J. \surnamestart Denning\surnameend}
  (\bibinfo{year}{1977}): \emph{\bibinfo{title}{Certification of Programs for
  Secure Information Flow}}.
\newblock {\sl \bibinfo{journal}{Commun. ACM}}
  \bibinfo{volume}{20}(\bibinfo{number}{7}), pp. \bibinfo{pages}{504--513},
  \doi{10.1145/359636.359712}.

\bibitemdeclare{article}{Enck2014}
\bibitem{Enck2014}
\bibinfo{author}{William \surnamestart Enck\surnameend}, \bibinfo{author}{Peter
  \surnamestart Gilbert\surnameend}, \bibinfo{author}{Byung-Gon \surnamestart
  Chun\surnameend}, \bibinfo{author}{Landon~P. \surnamestart Cox\surnameend},
  \bibinfo{author}{Jaeyeon \surnamestart Jung\surnameend},
  \bibinfo{author}{Patrick \surnamestart McDaniel\surnameend} \&
  \bibinfo{author}{Anmol~N. \surnamestart Sheth\surnameend}
  (\bibinfo{year}{2014}): \emph{\bibinfo{title}{{TaintDroid: An
  Information-Flow Tracking System for Realtime Privacy Monitoring on
  Smartphones}}}.
\newblock {\sl \bibinfo{journal}{Communications of the ACM}},
  \doi{10.1145/2494522}.

\bibitemdeclare{inproceedings}{Erlingsson:1999:SES:335169.335201}
\bibitem{Erlingsson:1999:SES:335169.335201}
\bibinfo{author}{\'{U}lfar \surnamestart Erlingsson\surnameend} \&
  \bibinfo{author}{Fred~B. \surnamestart Schneider\surnameend}
  (\bibinfo{year}{2000}): \emph{\bibinfo{title}{SASI Enforcement of Security
  Policies: A Retrospective}}.
\newblock In: {\sl \bibinfo{booktitle}{Proceedings of the 1999 Workshop on New
  Security Paradigms}}, \bibinfo{series}{NSPW '99}, \bibinfo{publisher}{ACM},
  \bibinfo{address}{New York, NY, USA}, pp. \bibinfo{pages}{87--95},
  \doi{10.1145/335169.335201}.
\newblock \urlprefix\url{http://doi.acm.org/10.1145/335169.335201}.

\bibitemdeclare{misc}{Ferreira2012}
\bibitem{Ferreira2012}
\bibinfo{author}{Paulo Jorge Abreu~Duarte \surnamestart Ferreira\surnameend}
  (\bibinfo{year}{2012}): \emph{\bibinfo{title}{MSc Dissertation. Information
  flow analysis using data-dependent logical propositions.}}
\newblock \bibinfo{note}{Faculdade de Ciências e Tecnologia, Universidade Nova
  de Lisboa}.

\bibitemdeclare{inproceedings}{tooljoana2013atps}
\bibitem{tooljoana2013atps}
\bibinfo{author}{J{\"u}rgen \surnamestart Graf\surnameend},
  \bibinfo{author}{Martin \surnamestart Hecker\surnameend} \&
  \bibinfo{author}{Martin \surnamestart Mohr\surnameend}
  (\bibinfo{year}{2013}): \emph{\bibinfo{title}{Using JOANA for Information
  Flow Control in Java Programs - A Practical Guide}}.
\newblock In: {\sl \bibinfo{booktitle}{Proceedings of the 6th Working
  Conference on Programming Languages (ATPS'13)}}.

\bibitemdeclare{article}{Lourenco2015}
\bibitem{Lourenco2015}
\bibinfo{author}{Lu\'{\i}sa \surnamestart Louren\c{c}o\surnameend} \&
  \bibinfo{author}{Lu\'{\i}s \surnamestart Caires\surnameend}
  (\bibinfo{year}{2015}): \emph{\bibinfo{title}{Dependent Information Flow
  Types}}.
\newblock {\sl \bibinfo{journal}{SIGPLAN Not.}}
  \bibinfo{volume}{50}(\bibinfo{number}{1}), pp. \bibinfo{pages}{317--328},
  \doi{10.1145/2775051.2676994}.

\bibitemdeclare{phdthesis}{Lourenco2016}
\bibitem{Lourenco2016}
\bibinfo{author}{Maria Luísa Sobreira~Gouveia \surnamestart
  Lourenço\surnameend} (\bibinfo{year}{2016}): \emph{\bibinfo{title}{A type
  system for value-dependent information flow analysis}}.
\newblock Ph.D. thesis.

\bibitemdeclare{inproceedings}{Myers2003}
\bibitem{Myers2003}
\bibinfo{author}{Andrew~C. \surnamestart Myers\surnameend} \&
  \bibinfo{author}{Barbara \surnamestart Liskov\surnameend}
  (\bibinfo{year}{2003}): \emph{\bibinfo{title}{{Protecting privacy using the
  decentralized label model}}}.
\newblock In: {\sl \bibinfo{booktitle}{Foundations of Intrusion Tolerant
  Systems, OASIS 2003}}, pp. \bibinfo{pages}{89--116},
  \doi{10.1145/363516.363526}.

\bibitemdeclare{}{JIF}
\bibitem{JIF}
\bibinfo{author}{Andrew~C. \surnamestart Myers\surnameend},
  \bibinfo{author}{Lantian \surnamestart Zheng\surnameend},
  \bibinfo{author}{Steve \surnamestart Zdancewic\surnameend},
  \bibinfo{author}{Stephen \surnamestart Chong\surnameend} \&
  \bibinfo{author}{Nathaniel \surnamestart Nystrom\surnameend}
  (\bibinfo{year}{2006}): \emph{\bibinfo{title}{Jif 3.0: Java information
  flow}}.
\newblock \urlprefix\url{http://www.cs.cornell.edu/jif}.

\bibitemdeclare{article}{Sabelfeld2003}
\bibitem{Sabelfeld2003}
\bibinfo{author}{Andrei \surnamestart Sabelfeld\surnameend} \&
  \bibinfo{author}{Andrew~C. \surnamestart Myers\surnameend}
  (\bibinfo{year}{2003}): \emph{\bibinfo{title}{{Language-based
  information-flow security}}}.
\newblock {\sl \bibinfo{journal}{IEEE Journal on Selected Areas in
  Communications}} \bibinfo{volume}{21}(\bibinfo{number}{1}), pp.
  \bibinfo{pages}{5--19}, \doi{10.1109/JSAC.2002.806121}.

\bibitemdeclare{misc}{LBS2001}
\bibitem{LBS2001}
\bibinfo{author}{Fred \surnamestart Schneider\surnameend},
  \bibinfo{author}{Greg \surnamestart Morrisett\surnameend} \&
  \bibinfo{author}{Robert \surnamestart Harper\surnameend}
  (\bibinfo{year}{2001}): \emph{\bibinfo{title}{{A Language-Based Approach to
  Security}}}.

\bibitemdeclare{}{FlowCaml}
\bibitem{FlowCaml}
\bibinfo{author}{V.~\surnamestart Simonet\surnameend} (\bibinfo{year}{2003}):
  \emph{\bibinfo{title}{The Flow Caml System (version 1.00): Documentation and
  user's manual}}.
\newblock
  \urlprefix\url{http://www.normalesup.org/~simonet/soft/flowcaml/manual/}.

\bibitemdeclare{article}{Vallee-Rai1999}
\bibitem{Vallee-Rai1999}
\bibinfo{author}{R~\surnamestart Vall{\'{e}}e-Rai\surnameend},
  \bibinfo{author}{P~\surnamestart Co\surnameend} \&
  \bibinfo{author}{E~\surnamestart Gagnon\surnameend} (\bibinfo{year}{1999}):
  \emph{\bibinfo{title}{{Soot-a Java bytecode optimization framework}}}.
\newblock {\sl \bibinfo{journal}{CASCON}}.

\bibitemdeclare{phdthesis}{ZdancewicPHD}
\bibitem{ZdancewicPHD}
\bibinfo{author}{Stephan~Arthur \surnamestart Zdancewic\surnameend}
  (\bibinfo{year}{2002}): \emph{\bibinfo{title}{Programming Languages for
  Information Security}}.
\newblock Ph.D. thesis, \bibinfo{address}{Ithaca, NY, USA}.
\newblock \bibinfo{note}{AAI3063751}.

\bibitemdeclare{article}{Zdancewic2004}
\bibitem{Zdancewic2004}
\bibinfo{author}{Steve \surnamestart Zdancewic\surnameend}
  (\bibinfo{year}{2004}): \emph{\bibinfo{title}{{Challenges for
  information-flow security}}}.
\newblock {\sl \bibinfo{journal}{Proceedings of the 1st International Workshop
  on the Programming Language Interference and Dependence (PLID'04)}}.

\bibitemdeclare{article}{Zhao2013}
\bibitem{Zhao2013}
\bibinfo{author}{Jianzhou \surnamestart Zhao\surnameend},
  \bibinfo{author}{Santosh \surnamestart Nagarakatte\surnameend},
  \bibinfo{author}{Milo~M.K. \surnamestart Martin\surnameend} \&
  \bibinfo{author}{Steve \surnamestart Zdancewic\surnameend}
  (\bibinfo{year}{2013}): \emph{\bibinfo{title}{Formal Verification of
  SSA-based Optimizations for LLVM}}.
\newblock {\sl \bibinfo{journal}{SIGPLAN Not.}}
  \bibinfo{volume}{48}(\bibinfo{number}{6}), pp. \bibinfo{pages}{175--186},
  \doi{10.1145/2499370.2462164}.

\end{thebibliography}
\end{document}